\documentstyle[preprint,aps,epsfig]{revtex}

\tightenlines

\begin{document}

\preprint{BNL-NT-00/27}

\title{Open and Hidden Charm Production
\\ in Heavy Ion Collisions at Ultrarelativistic Energies}
\author{
M.I. Gorenstein$^{a,b}$,
A.P. Kostyuk$^{a,b}$,
L. McLerran$^{c}$,
H. St\"ocker$^{a}$
 and
W. Greiner$^{a}$
}

\vspace{0.3cm}
\address{
$^a$ Institut f\"ur Theoretische Physik, Goethe Universit\"at,
Frankfurt am Main, Germany}

\address{$^b$ Bogolyubov Institute for Theoretical Physics,
Kyiv, Ukraine}

\address{$^c$ Physics Department,
Brookhaven National Laboratory, Upton, NY 11979 USA.}

\maketitle

\begin{abstract}
We consider the production of
the open charm and
$J/\psi$ mesons in heavy ion collisions at BNL RHIC.
We discuss several recently developed pictures for $J/\psi$ production and
argue that a measurement at RHIC energies is crucial for disentangling
these different descriptions.
\end{abstract}


\section{Introduction}

The production of charmonium states $J/\psi$ and $\psi^{\prime}$
have been measured in  nucleus--nucleus (A+A)
collisions at CERN SPS over the last 15 years by the NA38 and NA50
Collaborations.
This experimental program was motivated by a suggestion
of Matsui and Satz \cite{Satz1} to use the $J/\psi$
as a probe of the state of matter created in the early stage
of the collision (see also Ref.\cite{Satzr} for a recent review).

In the original picture \cite{Satz1}, it is assumed that
$J/\psi$ particles are created in primary
nucleon--nucleon (N+N) collisions when the charm-anticharm
quark pairs are produced.
The fraction (a few percents) of initial charm pairs which
formed a
$J/\psi$ state is subsequently reduced in A+A collisions because of
the $J/\psi$ suppression by a factor which depends upon the details
associated with the impact parameter and energy of the collisions.

The statistical approach, proposed in Ref.\cite{Ga1}, assumes that
$J/\psi$ mesons are created at hadronization according to
the available hadronic phase-space similar to other (lighter) hadrons.
We call this a hadron gas (HG) model.

Recently a picture of the $J/\psi$ creation via
coalescence (recombination)
of charmed quarks was developed within several model approaches
\cite{Kabana:2000sd,Br1,Go:00,Le:00,Ra:00}.
Similar to the HG model \cite{Ga1},
the charmonium states are assumed
to be formed at the hadronization stage.
However, they are produced as a coalescence of  earlier
created $c$ and $\overline{c}$ quarks. Therefore,
the $J/\psi$ yield and the total charm yield are connected
to each other and differ in
general from the HG equilibrium values.

A principal quantity to be studied in our paper is the ratio
of the the $J/\psi$ multiplicity to the total number
of $c\overline{c}$ pairs, $N_{c\overline{c}}$, created in A+A collisions:
\begin{equation}\label{ratio}
R~\equiv ~\frac{\langle J/\psi \rangle } {N_{c\overline{c}}}~.
\end{equation}
In the standard picture of $J/\psi$ suppression, the number of primary
$J/\psi$ particles is
proportional to the number of directly produced $c\overline{c}$ pairs
with invariant mass below $D\overline{D}$ meson threshold
\cite{Gavai}. The fraction of the subthreshold pairs in the total number
of $N_{c\overline{c}}$ decreases when the collision energy
$\sqrt{s}$ grows.
The number of initially produced $J/\psi$ particles is further
reduced because of their inelastic interactions with nucleons from the
colliding nuclei.
This effect becomes stronger when the number of
participants increases. In addition, $J/\psi$ particles can be destroyed
by the secondary hadrons (co-movers) created in the A+A collision.
The density of these hadrons increases with both $N_p$
and $\sqrt{s}$.
The decreasing of the ratio $R$ due to the above
mentioned effects is known as {\it `normal $J/\psi$ suppression'}.
The formation of quark-gluon plasma (QGP)
at large values of $\sqrt{s}$ and $N_p$ is supposed to
lead to a sudden and strong drop of the ratio (\ref{ratio}). This
is called {\it `anomalous $J/\psi$ suppression'}.
Therefore, there is an unambiguous consequence of the standard
picture \cite{Satz1,Satzr}:
when the collision energy per nucleon pair
$\sqrt{s}$ and/or the  number of nucleon
participants $N_p$  increase, the ratio (\ref{ratio})
decreases.

In models where the $J/\psi$ mesons are produced by recombination
of charm quarks at the hadronization stage, the
picture can be much different.
For the large number of $c\overline{c}$ pairs $N_{c\overline{c}}>>1$
expected at the RHIC energies, the
charmonium states can be formed from a
$c$ and $\overline{c}$ which were initially created in
different $c\overline{c}$ pairs \cite{Ra:00}. Therefore, the
multiplicity of $J/\psi$ due to recombination of charm quarks
should be roughly proportional to:
\begin{equation}\label{recom}
\langle J/\psi \rangle ~\sim
\frac{\left(N_{c\overline{c}}\right)^2}{N_{tot}}~,
\end{equation}
where $N_{tot}$ is the total multiplicity of produced
hadrons. In this case  the ratio (\ref{ratio})
may be constant or even increase
(we call this {\it `$J/\psi$ enhancement'})
with $N_p$ and/or $\sqrt{s}$
depending on the underlying physics which generates
$c\overline{c}$ pairs and charmonium states.

We consider the canonical ensemble 
formulation \cite{Go:00}  of the statistical coalescence model (SCM)
\cite{Br1} to describe the
$c$ and $\overline{c}$ recombination into the
charmonium states at the hadronization stage.  A canonical rather than the
more familiar grand canonical formalism is required as we must properly
account for the small numbers of charm quarks which are present.  When
large numbers of charm quarks are produced in each collision, these two
formalisms become equivalent, but the grand canonical formalism fails
for small numbers.
As in the standard picture, the charm is
created at the very early stage of A+A reaction via hard parton
collisions.
The $J/\psi$ formation is however very different in the
standard picture of the $J/\psi$ suppression
\cite{Satz1} and in the SCM. This leads to very different relations
between the open charm and charmonium yields.

In the analysis below we present also two
possibilities of charm chemical
equilibration.  First, we assume that
both the open charm and charmonium yields correspond to
a chemically equilibrated hadronic gas.
This scenario is the most extreme in its assumptions, but provides a
benchmark for the dispersion of various theoretical results.
Another possibility is that
the number of $c\overline{c}$ pairs, which is an input for the SCM,
equals to the equilibrium value in the QGP just before its
hadronization.
These two possibilities are the extremes
of scenarios with chemical equilibration of charm.
We do not pretend that these extreme cases really
represent what happens in heavy ion collisions.
They are studied to complete our analysis.
In fact, the hard parton collisions seem to be the dominant source
of the $c\overline{c}$ pairs created at the RHIC energies.

The issue we want to address in this paper is how to disentangle the
different scenarios of charm and charmonium production.
It will be shown that the different pictures have a much
different dependence on both  $N_p$ and  $\sqrt{s}$.
Distinctive signatures of each
of these pictures will be found.
We shall
argue that $J/\psi$ measurements at RHIC energies, especially in
combination with a measurement of open charm, will go far towards
resolving the
various pictures.
Of course, to really disentangle
different scenarios, it is useful to be a little more quantitative.
That is the central purpose of this paper.

The outline of this paper is as follows.
In Section 2 we present the equilibrium  HG
model for the open charm and charmonium production.
The picture of
the SCM is described in
Section 3.  The results and conclusions
of the paper are presented respectively in Sections 4 and 5.

\section{Hadron Gas  Model}

The equilibrium HG model describes
the hadron yields measured in A+A collisions in terms
of three parameters:  volume $V$, temperature $T$
and baryonic chemical potential $\mu_B$.
This model seems to work in the AGS--SPS energy
region describing successfully not only the pion and (anti)nucleon
yields but (multi)strange hadron abundances as well (see e.g. \cite{HG}).
For the  RHIC energies the temperature parameter $T$ is expected to be
similar to that for the SPS energies: $T=175\pm 10$~MeV.
The baryonic chemical potential becomes small
 ($\mu_B < T$) and decreases with the collision energy.

Does the HG model work for the charm and charmonium yields
in A+A collisions?
The answer on this question is rather uncertain at present.
There are no open charm yield data measured in A+A
collisions. This may be a reason why there were no
attempts to calculate the open charm hadron yield within
the HG model.

\subsection{Grand Canonical Ensemble Treatment of the Hadron Gas}

The HG model assumes the
following formula for the hadron thermal multiplicities
in the grand canonical ensemble:
\begin{equation}\label{stat}
N_j~=~\frac{d_j~V}{2\pi^2} ~
\int_0^{\infty}p^2dp~\left[\exp\left(
\frac{\sqrt{p^2+m_{j}^2} - \mu_j}{T}\right)~\pm~1\right]^{-1}~,
\end{equation}
where $V$ and $T$ correspond to the volume\footnote{To avoid further
complications we use ideal HG formulas and neglect excluded volume
corrections.} and
temperature
of the HG system, $m_j$, $d_j$ denote hadron masses and
degeneracy
factors.
The particle chemical
potential $\mu_j$ in Eq.(\ref{stat})
is defined as
\begin{equation}\label{mui}
\mu_j~=~b_j\mu_B~+~s_j\mu_S~+~c_j\mu_C~,
\end{equation}
where $b_j,s_j,c_j$ denote the baryonic number strangeness and
charm of particle $j$. The baryonic chemical potential $\mu_B$
regulates the baryonic density of the HG system whereas
strange $\mu_S$ and charm $\mu_C$ chemical potentials should be found
from the requirement of zero value for the total strangeness and charm
in the system (in our consideration we neglect small effects
of a non-zero electrical chemical potential).

Some of the particles which are observed after a nuclear collision result from
the decay of resonances which have decoupled after the matter fell out
of equilibrium.
The total multiplicities $N_j^{tot}$ of such observed particles in the HG model
are therefore the sum of:
\begin{equation}\label{dec}
N_j^{tot}~=~N_j ~+~\sum_R {\rm Br}(R \rightarrow j) N_R~,
\end{equation}
where ${\rm Br}(R \rightarrow j)$ are the corresponding resonance decay
branching ratios (both strong and electromagnetic decays are taken into
account). The temperatures and chemical potentials used in these
relationships are those at decoupling. The hadron yield ratios
$N_j^{tot}/N_i^{tot}$ in the grand canonical ensemble are therefore known
functions of $T$ and $\mu_B$ variables, and are independent of the volume
parameter $V$.

For the thermal multiplicities of both open charm and
charmonium states the Bose and Fermi effects are negligible.
This is because the charm quark mass is so large that the system is always
dilute for charm quark densities.
Therefore, Eq.(\ref{stat}) is simplified to:
\begin{equation}\label{gce}
N_j~=~ \frac{d_j
V~e^{\mu_j/T}}{2\pi^2}~T~m^2_j~K_2\left(\frac{m_j}{T}\right)~
\simeq~d_j~V~e^{\mu_j/T}~\left(\frac{m_jT}{2\pi}\right)^{3/2}~\exp\left(-
\frac{m_j}{T}\right)~,
\end{equation}
where $K_2$ is the modified Bessel function.
The HG model
gives  the $J/\psi$ yield:
\begin{equation}\label{psitot}
N_{J/\psi}^{tot}=N_{J/\psi}~+~{\rm Br}(\psi^{\prime})N_{\psi^{\prime}}~+~
~{\rm Br}(\chi_1)N_{\chi_1}~+~
{\rm Br}(\chi_2)N_{\chi_2}~,
\end{equation}
where $N_{J/\psi}$, $N_{\psi^{\prime}}$, $N_{\chi_1}$, $N_{\chi_2}$ are
calculated according to Eq.(\ref{gce}) and ${\rm Br}(\psi^{\prime})\simeq
0.54$, ${\rm Br}(\chi_1)\simeq 0.27$, ${\rm Br}(\chi_2)\simeq 0.14$ are
the decay branching ratios of the excited charmonium states into $J/\psi$.
Note that $T\simeq170\div 180$~MeV leads to the HG value of the thermal
ratio of $\langle \psi^{\prime} \rangle /\langle J/\psi \rangle$:
\begin{equation}\label{psiratio}
 \frac{\langle \psi^{\prime}\rangle}
{\langle J/\psi \rangle} ~=~
\left(\frac{m_{\psi^{\prime}}}{m_{J/\psi}}\right)^{3/2}~
\exp \left(-~\frac{m_{\psi^{\prime}} - m_{J/\psi}}{T}\right)~=~0.04\div
0.05~,
\end{equation}
in agreement with data \cite{ratio} in central ($N_p>100$) Pb+Pb
collisions. This fact was first noticed in Ref.\cite{Sh}.
At small $N_p$ as well as in p+p and p+A collisions the
measured value of the  $\langle \psi^{\prime} \rangle /\langle J/\psi
\rangle$ ratio is several times larger than its statistical mechanics
estimate (\ref{psiratio}). Therefore, different versions of the
statistical approach considered in this paper will be restricted
to A+A collisions with $N_p> 100$. We do not intend here
to describe the open and hidden charm production in p+p, p+A
and in very peripheral A+A collisions within the statistical models.

Of course the temperature of decoupling of the hadron gas is quite large,
and one should properly be worried about the consistency of such a description.
It nevertheless does provide a fair description of available data on
particle production.  As such, it provides a standard against which
interesting new physics might be calibrated.

\subsection{The Canonical Ensemble for a Hadron Gas}

The numbers of charm quarks produced in high energy nuclear collisions
need not be large.  Implicit in the grand canonical ensemble treatment is this
large number assumption.  The canonical ensemble treatment is valid
for any small number of particles, and we now employ it to properly
describe charm abundances.

In the canonical ensemble formulation the thermal charmonium
multiplicities are still given by Eq.(\ref{gce}) for a grand canonical
ensemble as charmonium states have zero net charm. The multiplicities
given by this equation (\ref{gce}) for open charm hadrons are modified by
an additional multiplicative `canonical suppression' factor. This has been
shown, for example, in Reference \cite{Go1}). The suppression factor is
the same for all individual open charm states. Therefore, if $N_O$ is the
total grand canonical ensemble multiplicity of all open charm and
anticharm mesons and (anti)baryons, then the canonical ensemble (c.e.) value of
the total open charm is equal to:
\begin{equation}\label{NOce}
N^{c.e.}_{O}~=~N_O~\frac{I_1(N_O)}{I_0(N_O)}~,
\end{equation}
where $I_0,I_1$ are the modified Bessel functions. To find $N_O$ we use
Eq.(\ref{gce}) for thermal multiplicities of the open charm hadrons in the
grand canonical ensemble and take the summation over all known particles
and resonances with open charm \cite{pdg}. The factor $I_1(N_O)/I_0(N_O)$
in this equation is due to the exact charm conservation. For $N_O<<1$ one
has $I_1(N_O)/I_0(N_O)\simeq N_O/2$ and, therefore, the canonical ensemble
total open charm multiplicity $N_0^{c.e.} \sim N_0^2 << N_0 $ is strongly
suppressed in comparison to the corresponding grand canonical ensemble
result. At SPS energies the canonical ensemble suppression effects, due to
small numbers of produced charm, are important for the thermal open charm
yield even in the most central Pb+Pb collisions. These suppression effects
become crucial when the number of participants $N_p$ decreases. Note that
for $N_O << 1$ the multiplicities of the open charm hadrons are
proportional to $V^2$ in the canonical ensemble HG formulation (instead of
$V$ in the grand canonical ensemble). In the opposite case $N_O>>1$,
$I_1(N_O)/I_0(N_O)\rightarrow 1$, therefore the grand canonical ensemble
and canonical ensemble results coincide: $N^{c.e.}_{O}~\rightarrow~N_O$.
In a high energy collision of heavy nuclei, the average numbers of
produced (anti)baryons, (anti)strange hadrons and charged particles are
much larger than $1$. Due to this fact, the baryonic number, strangeness
and electric charge of the HG system can be treated according to the grand
canonical ensemble. This is not true in general for charm.

The number of $c\overline{c}$ pairs in the HG,
$N^{HG}_{c\overline{c}}$, is given by
\begin{equation}\label{NccHG}
N^{HG}_{c\overline{c}}~=~\frac{1}{2}~
N_O~\frac{I_1(N_O)}{I_0(N_O)}~
+~N_{H}~,
\end{equation}
where $N_{H}$
is the total HG
multiplicity of particles with hidden charm.

We can now calculate the $J/\psi$ multiplicity, since this involves only
knowing the grand canonical ensemble results for $c \overline c$ bound states,
and the total open charm multiplicity $N^{c.e.}_{O}$, since we now have
the canonical ensemble results in hand
(\ref{NOce}).  We need
the chemical freeze-out parameters $V,T,\mu_B$ for A+A collisions.
For the  RHIC energies the temperature parameter $T$ is expected to be
similar to that for the SPS energies: $T=175\pm 10$~MeV.
To fix the unknown
volume parameter $V$ and baryonic chemical
potential $\mu_B$ we use the parametrization of the total
pion multiplicity \cite{Ga:pi}:
\begin{equation}\label{pionexp}
\frac{\langle \pi \rangle}{N_p} ~\simeq ~C~
\frac{(\sqrt{s}-2m_N)^{3/4}}{(\sqrt{s})^{1/4}},
\end{equation}
where $C \approx 1.46$~GeV$^{-1/2}$ and  $m_N$
is the nucleon mass.
Eq.(\ref{pionexp}) is in an agreement with both the SPS data and the
preliminary RHIC data in Au+Au collisions at
$\sqrt{s}=56$~GeV and $ \sqrt{s}=130$~GeV.
The pion multiplicity (\ref{pionexp}) should be equated to
the total HG pion multiplicity $N_{\pi}^{tot}$
calculated according to Eq.(\ref{dec}) which accounts for resonance decays
into pionic final states.
The HG parameters $V$ and $\mu_B$ are found then as the solution
of the following coupled equations:
\begin{eqnarray}
\label{pi}
\langle \pi \rangle ~ & = & N_{\pi}^{tot}(V,T,\mu_B)~\equiv~
V~n_{\pi}^{tot}(T,\mu_B)~,\\
 \label{Np}
N_p~ & = & ~V~n_B(T,\mu_B)~,
\end{eqnarray}
where $n_B$ is the HG baryonic density. In
these calculations we fix the temperature parameter $T$.

The baryonic chemical potential for Au+Au collisions
at the RHIC energies is small
 ($\mu_B < T$) and decreases with collision energy. Therefore,
most of the thermal HG multiplicities
become close to their limiting  values at
 $\mu_B \rightarrow 0$.
Consequently most of hadron ratios $N_j^{tot}/N_i^{tot}$
become independent of the collision energy.

If we assume that the charm quarks reach their chemically equilibrated values,
then the chemical potential for charm vanishes.  We can then compute the
abundances of charmonium.
The ratio of $J/\psi$ to
negatively charged
hadrons is only weakly dependent on the collision energy as a consequence
of the considerations of the previous paragraph
and  equals approximately to:
\begin{equation}\label{psihmin}
\frac{N_{J/\psi}^{tot}}{\langle h^- \rangle}~\simeq ~(0.6 \div 1.5)~10^{-6}~,
\end{equation}
for $T=170$~MeV and $T=185$~MeV, respectively.
The result (\ref{psihmin}) is close to the estimate
\cite{Ga1,GaJ}  of this ratio, $1.0~10^{-6}$, extracted
from the NA50 data in Pb+Pb collisions at 158 A~GeV.

\subsection{Summary of Results}

In this section, we developed the grand canonical ensemble and canonical
ensemble formalism for describing abundances of particles in thermal
equilibrium.  Nucleon and pion abundances were used to constrain the
parameters of these treatments: volume and baryon
chemical potential. The temperature of the chemical freeze-out was assumed to be 
equal to its value at SPS energies.
Formulae were derived for the abundances of open
charm and hidden charm ($c \overline c$ states).  In particular, the
charm to negative charge particle multiplicity comes out
correctly within such a naive treatment.

\section{Statistical Coalescence Model}

One has serious reservation about the concept of charm
chemical equilibration within HG approach.   The HG system is
expected to exist for a much shorter time than is needed for the
chemical equilibration of charm. It is more reasonable to assume
that the charm quark-antiquark pairs are created at the early
stage of A+A reaction and the number of $c\overline{c}$
pairs remains approximately unchanged during the further evolution.
After the hadronization, the system
would correspond then to the HG picture with a nonequilibrium
value of the total charm.

We imagine two possible scenarios.  In the first, we imagine the
$c \overline c$ pairs reach their equilibrium values in the quark gluon
plasma which precedes the hadronic gas.  We call this the QGP + SCM model,
where the SCM means statistical coalescence model.  In the second, we imagine
the $c \overline c$ pairs are produced by hard QCD processes early in the
collision, and are unchanged by the presence of the hadronic gas or of
the quark gluon plasma.  This scenario we refer to as the pQCD + SCM model.

\subsection{Charm Quarks Not In Chemical Equilibrium}

The SCM
\cite{Br1,Go:00} assumes that the charmonium states
are formed at the hadronization stage.
This is similar to the HG model \cite{Ga1}.
However, the  charmonium states are produced via a coalescence of
earlier created
$c\overline{c}$ quarks.
The number of $c\overline{c}$
pairs, $N_{c\overline{c}}$, differs in general from the result
expected in the  equilibrium HG.
One needs then a new parameter
 $\gamma_c$  \cite{Br1} to
adjust the thermal HG results to the required number of
$N_{c\overline{c}}$ fixed by the early stage evolution. In the grand
canonical ensemble, $\gamma_c = e^{\mu/T}$, where $\mu$ is the chemical
potential for $c + \overline c$ pairs. This is analogous to the
introduction of the strangeness suppression factor $\gamma_s$ \cite{Raf1}
in the HG model, when the total strangeness observed is smaller than its
chemical equilibrium value.

At high collision energies we find $\gamma_c>1$ so that the open charm
hadron yield is enhanced by a factor $\gamma_c$ and charmonium yield by a
factor $\gamma_c^2$ in comparison with the equilibrium HG predictions. The
canonical ensemble formulation of the SCM, as shown in Ref. \cite{Go:00},
gives
\begin{equation}\label{Ncc1}
N_{c\overline{c}}~=~\frac{1}{2}~
\gamma_c~N_O~\frac{I_1(\gamma_c N_O)}{I_0(\gamma_cN_O)}~
+~\gamma_c^2~N_{H}~.
\end{equation}
Note that the second term in the right-hand side of this equation is
proportional to the number of hidden charm states, $N_H$, and is typically
very small compared to the total number of charm quarks. It gives only a
tiny correction to the first term. Most of the created $c\overline{c}$
pairs are transformed into the open charm hadrons.

If $N_{c\overline{c}} >> 1$, Eq.(\ref{Ncc1})
is simplified to \cite{Br1}:
$N_{c\overline{c}}=\gamma_c~N_O/2+\gamma_c^2N_{H}$.
This happens for central  Au+Au collisions
at upper RHIC energy  $\sqrt{s}=200$~GeV.
For  lower RHIC energy $\sqrt{s}=56$~GeV
the value of
$N_{c\overline{c}}$
could be close to (or even smaller than) unity so that
the canonical ensemble suppression effects for the open charm are
still important. Note that for the non-central A+A collisions
the total charm decreases essentially so that the
canonical ensemble suppression
effects become stronger,
hence their consideration is  necessary to study the
$N_p$ dependence of the charmonium production even
at the upper RHIC energy.

The above equation for the number of $c \overline c$ pairs
will be used to find the charm enhancement
factor $\gamma_c$ and calculate  the $J/\psi$ multiplicity:
\begin{equation}\label{Npsi}
\langle J/\psi \rangle~= ~\gamma_c^2~N_{J/\psi}^{tot}~.
\end{equation}
where $N_{J/\psi}^{tot}$ is given by Eq.(\ref{psitot}), the equation which
accounts for decays of states with hidden charm into the $J/\psi$.

\subsection{Estimating the Initial Charm Production}

The number of
$c\overline{c}$ pairs, $N_{c\overline{c}}$, in the left-hand
side of Eq.(\ref{Ncc1}) will be estimated in
two different ways -- in the pQCD approach,
$N_{c\overline{c}}=N^{pQCD}_{c\overline{c}}$, and in
the equilibrium QGP just before its hadronization,
$N_{c\overline{c}}=N^{QGP}_{c\overline{c}}$,  --
and  used then as the input for the SCM.

\vspace{0.3cm}
The pQCD calculations for ${c\overline{c}}$ production
cross section $\sigma(pp\rightarrow
c\overline{c})$ in p+p collisions
were first done in Ref.\cite{comb}.
In Ref.\cite{ruusk} the cross section
$\sigma(pp\rightarrow
c\overline{c})$
was calculated in the next-to-leading order pQCD approximation
and  fit  to the existing p+p data for charm production.
We have parametrized the $\sqrt{s}$-dependence
of the charm production cross
section at $10 < \sqrt{s} < 200$~GeV
as:
\begin{equation}\label{pert1}
\sigma(pp\rightarrow c\overline{c})~=~\sigma_0~
\left(1- \frac{M_{0}}{\sqrt{s}}\right)^{\alpha}~
\left(\frac{\sqrt{s}}{M_{0}}\right)^{\beta}~,
\end{equation}
with $\sigma_0 \simeq 3.392 $~$\mu$b, $M_{0}\simeq 2.984$~GeV,
$\alpha \simeq 8.185$ and $\beta \simeq 1.132$.
For our fit we used the result of Ref.\cite{ruusk} obtained with
GRV HO set of proton structure functions \cite{GRV} for charm
quark mass and renormalization scale $m_c=\mu=1.3$~GeV.
The formula (\ref{pert1}) agrees approximately with existing
data and leads to the value of
$\sigma(pp\rightarrow
c\overline{c})~\simeq 0.35$~mb at $\sqrt{s}=200$~GeV.

The number of produced $c\overline{c}$ pairs in A+A collisions
is proportional to the number of primary nucleon-nucleon collisions:
\begin{equation}\label{pertNp}
N^{pQCD}_{c\overline{c}}~=~ N_{coll}^{AA}(N_p)~
\frac{\sigma(pp\rightarrow c\overline{c})}{\sigma_{NN}^{inel}}~,
\end{equation}
where $N_{coll}^{AA}(N_p)$ is the number of primary nucleon-nucleon
collisions, which depends on the number of participants and on
the details of nuclear geometry,
$\sigma_{NN}^{inel} \approx 30~$mb is the inelastic nucleon-nucleon cross
sections.

For the most central A+A collision ($N_p \simeq 2A$) the number
of primary nucleon collisions can be estimated
within the model of homogeneous spherical nucleus \cite{eskola}:
\begin{equation}\label{Ncoll}
N_{coll}^{AA}(N_p)~\simeq ~\frac{9}{8} \frac{A^2}{\pi
R_A^2}~\sigma_{NN}^{inel},
\end{equation}
with $R_A \approx 1.12 A^{1/3}$~fm. Therefore $N_p$ dependence on the
number of produced $c\overline{c}$ pairs can be calculated
as:\footnote{Strictly speaking, Eq.(\ref{pert}) does not exactly
correspond to the real experimental situation. $Au+Au$ collisions with
different centrality rather than central collisions of nuclei with
different atomic numbers will be studied at RHIC. Still, the formula
(\ref{pert}) gives almost the same numerical result as the more realistic
calculations \cite{suprenh}.}
\begin{equation}\label{pert}
N^{pQCD}_{c\overline{c}}~\simeq ~C~
\sigma(pp\rightarrow c\overline{c})~N_p^{4/3}~,
\end{equation}
where $C \approx 11$ barn$^{-1}$.

\vspace{0.3cm}

\subsection{Initial Charm from a Chemically Equilibrated QGP}

We also make the SCM model calculations
with $N_{c\overline{c}}=N^{QGP}_{c\overline{c}}$
in the left-hand side
of Eq.(\ref{Ncc1}). The
values of
$N^{QGP}_{c\overline{c}}$ in central Au+Au
collisions at different $\sqrt{s}$ are calculated as the equilibrium
number of $c\overline{c}$ pairs in the QGP just before its
hadronization.
The hadronization temperature of the QGP is taken to be equal
to the chemical freeze-out temperature in the HG, i.e.
we assume that the chemical freeze-out in the HG takes
place just after the QGP hadronization.
The volume $V_{QGP}$
of the QGP is found then from the requirement
of an approximate entropy conservation during
the hadronization transition: $S_{QGP}=S_{HG}$
(the nonzero but small baryonic chemical
potentials in the HG and in the QGP do not affect the QGP volume
estimate).
The calculation of $N^{QGP}_{c\overline{c}}$ corresponds
to the equilibrium
ideal gas of quarks and
gluons ($m_u\approx m_d\approx 0$, $m_s \approx 0.12 $~GeV,
$m_c \approx 1.3$~GeV).
In the grand canonical ensemble,
one finds for the number of $c$ and $\overline{c}$:
\begin{equation}\label{NQgce}
N_{c}~=~N_{\overline{c}}~=~ \frac{d_c~V_{QGP}}
{2\pi^2}~T~m^2_c~K_2\left(\frac{m_c}{T}\right)~,
\end{equation}
where $d_c=6$ is the degeneracy factor for the $c$-quark.
Taking into account the exact charm conservation
we obtain the number of $c\overline{c}$ pairs in the canonical ensemble:
\begin{equation}
\label{NQce}
N^{QGP}_{c\overline{c}}~=~\frac{1}{2}~
\left(N_{c} +N_{\overline{c}}\right)~
\frac{I_1\left(N_{c} +N_{\overline{c}}\right)}
{I_0\left(N_{c} +N_{\overline{c}}\right)}~.
\end{equation}

\section{Comparison of Various Mechanisms}

In this Section we present the results of the SCM and the HG model.
Two different inputs
for $N_{c\overline{c}}$
in the SCM (\ref{Ncc1}) are studied:
$N^{pQCD}_{c\overline{c}}$ (\ref{pert}) and
$N^{QGP}_{c\overline{c}}$ (\ref{NQce}). We call these model
formulations as pQCD+SCM and QGP+SCM, respectively.
The HG model, discussed in Section 2, can be considered
as a special case of the SCM when $\gamma_c \equiv 1$.
First, we consider central Au+Au collisions ($N_p\simeq 2$A) at different
collision energies $\sqrt{s}$. Keeping the temperature parameter $T$
fixed we solve Eqs.(\ref{pi},\ref{Np}) to find
the hadronization parameters $V$ and $\mu_B$.
Eq.(\ref{pionexp}) is used as the parametrization of the
 $\sqrt{s}$ dependence for the total pion
multiplicity.
With HG parameters $T$, $V$ and $\mu_B$ we calculate
the thermal multiplicities of individual open and hidden
charm states in the grand canonical ensemble according to Eq.(\ref{gce})
and then find the grand canonical ensemble total open charm $N_O$ and
the total hidden charm  $N_H$ multiplicities taking
the summation over all known open and hidden charm states.
Using the calculated values of $N_O$ and $N_H$ we proceed
further with three different model
formulations along the scheme presented in Table 1.

\subsection{Number of $c \overline c$ Pairs as Function of Energy}

\begin{figure}[t]
\begin{center}
\vfill
\leavevmode
\epsfysize=17.5cm \epsfbox{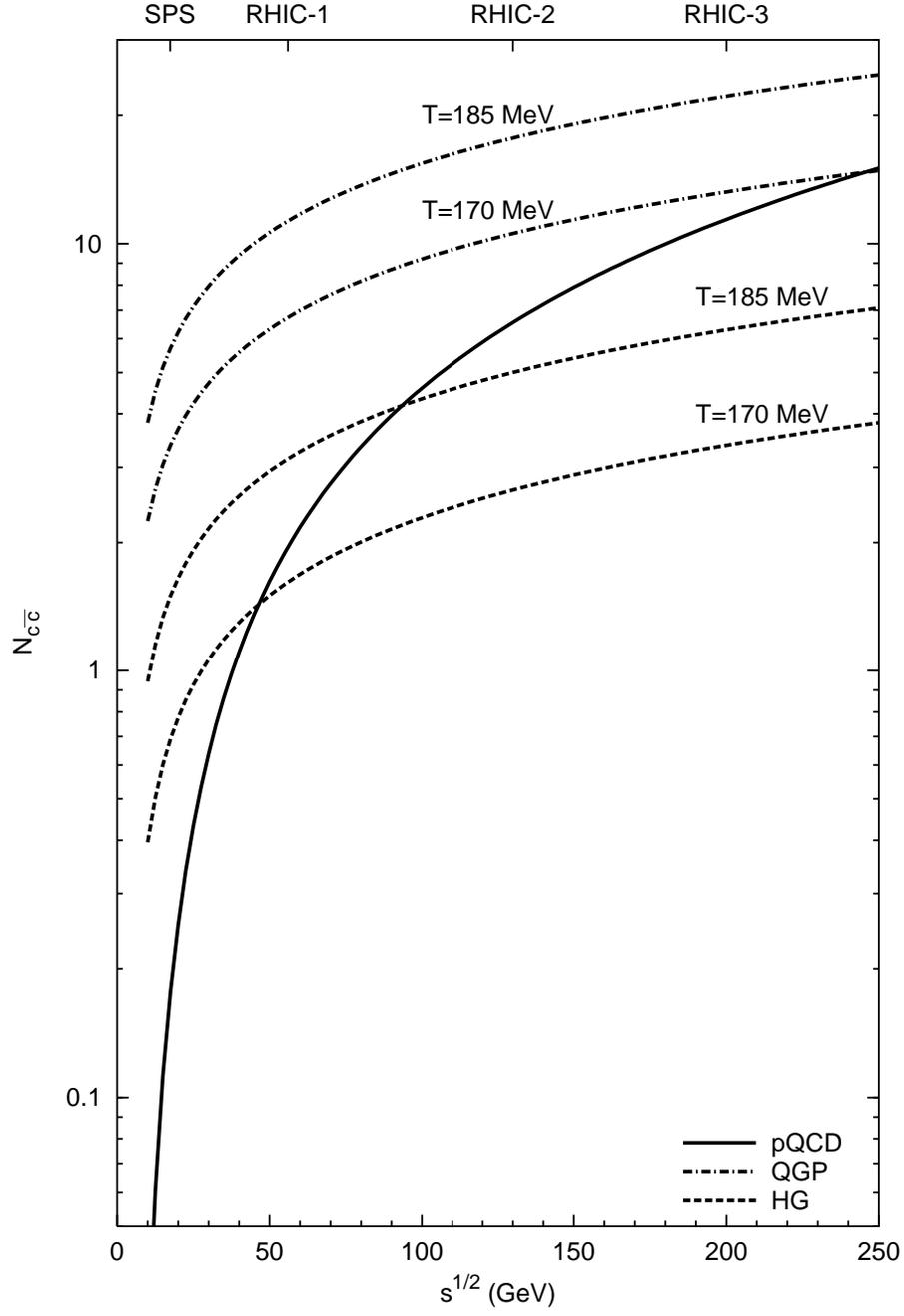}
\vfill
\mbox{}\\
\caption{The dependence of the average number of $c\overline{c}$ pairs per
Au+Au central collision
on the center-of-mass energy of the nucleon pair.
The calculations are done in
the framework of three
different models: perturbative QCD (pQCD), equilibrium quark-gluon
plasma (QGP), equilibrium hadron gas (HG). The QGP and HG results are shown
for two different temperatures: $T=170$ MeV and $T=185$ MeV.
\label{Ncc}
}
\end{center}
\end{figure}

The dependence of the number of $c\overline{c}$ pairs,
$N_{c\overline{c}}$, on the collision energy
is shown in Fig.~\ref{Ncc} for different model formulations.
The value of $N^{pQCD}_{c\overline{c}}$ (\ref{pert}) increases strongly
with collision energy and at the highest RHIC energy it becomes
several times larger than the thermal equilibrium value
$N^{HG}_{c\overline{c}}$ in the HG model (\ref{NccHG}).
At the SPS energies $\sqrt{s}\approx 20$~GeV
the situation is however different: $N^{HG}_{c\overline{c}} >
N^{pQCD}_{c\overline{c}}$.
The value of $N^{QGP}_{c\overline{c}}$ defined by Eq.~(\ref{NQce})
seems to give an upper limit of the charm production in
A+A collisions. The inequality $N^{QGP}_{c\overline{c}}>>
N^{HG}_{c\overline{c}}$ is mainly because of a small
mass of charm quark, $m_c\simeq 1.3$~GeV, in comparison to
the masses of charmed hadrons ($D$, $D^*$, etc.).

\subsection{$J/\psi$ Multiplicity versus Energy}

\begin{figure}[t]
\begin{center}
\vfill
\leavevmode
\epsfysize=17.5cm \epsfbox{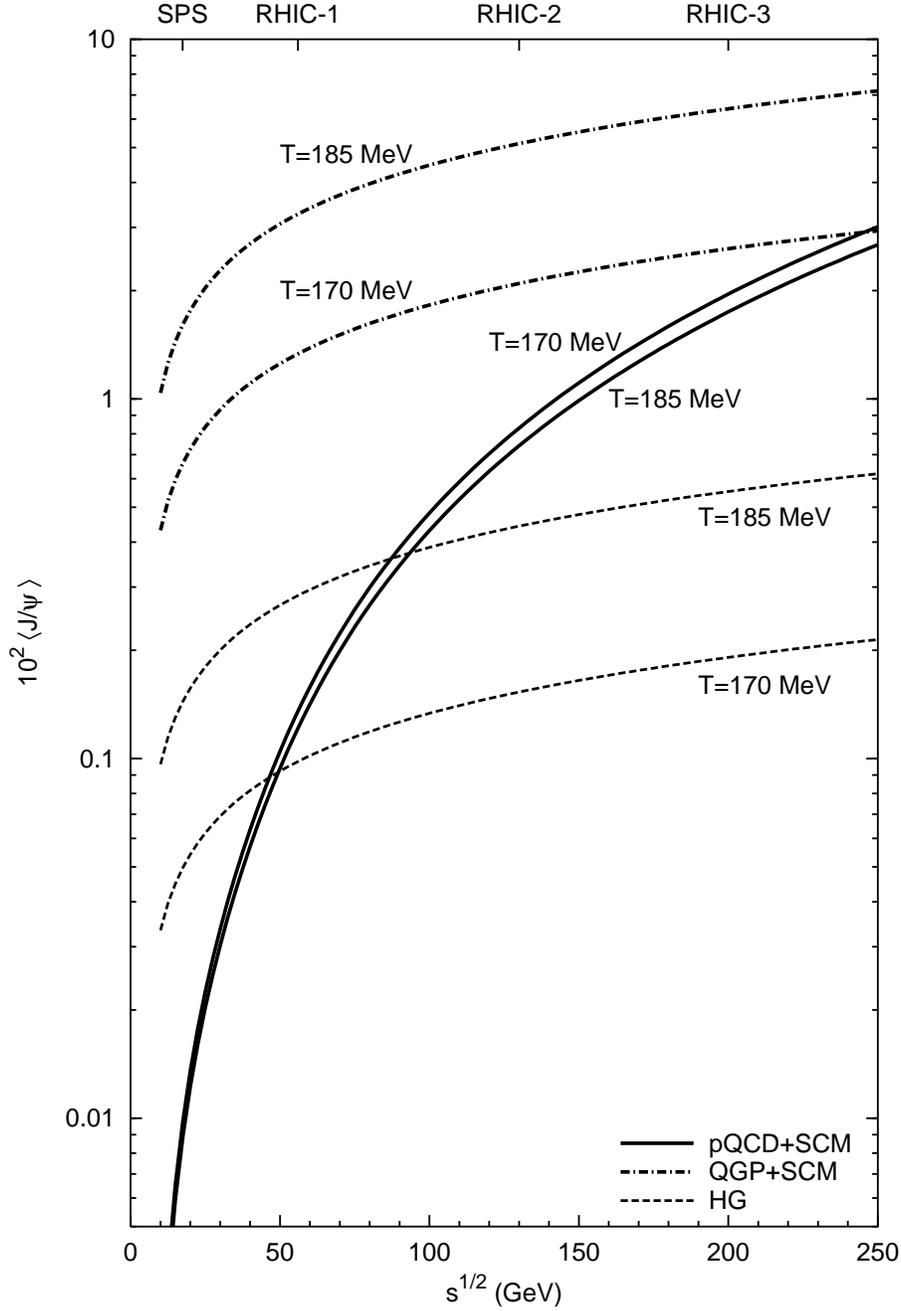}
\vfill
\mbox{}\\
\caption{The dependence of the average number of $J/\psi$-particles per
Au+Au central collision
on the center-of-mass energy of the nucleon pair.
The calculations are done in
the framework of three
different models: statistical coalescence
model with the number of
$c\overline{c}$ pairs given by perturbative QCD (pQCD+SCM), statistical
coalescence model with the number of $c\overline{c}$ pairs given by
equilibrium quark-gluon plasma (QGP+SCM) and equilibrium hadron gas model
(HG). All the dependences are shown
for two different temperatures: $T=170$ MeV and $T=185$ MeV.
\label{Jpsi}
}
\end{center}
\end{figure}

\begin{figure}[t]
\begin{center}
\vfill
\leavevmode
\epsfysize=17.5cm \epsfbox{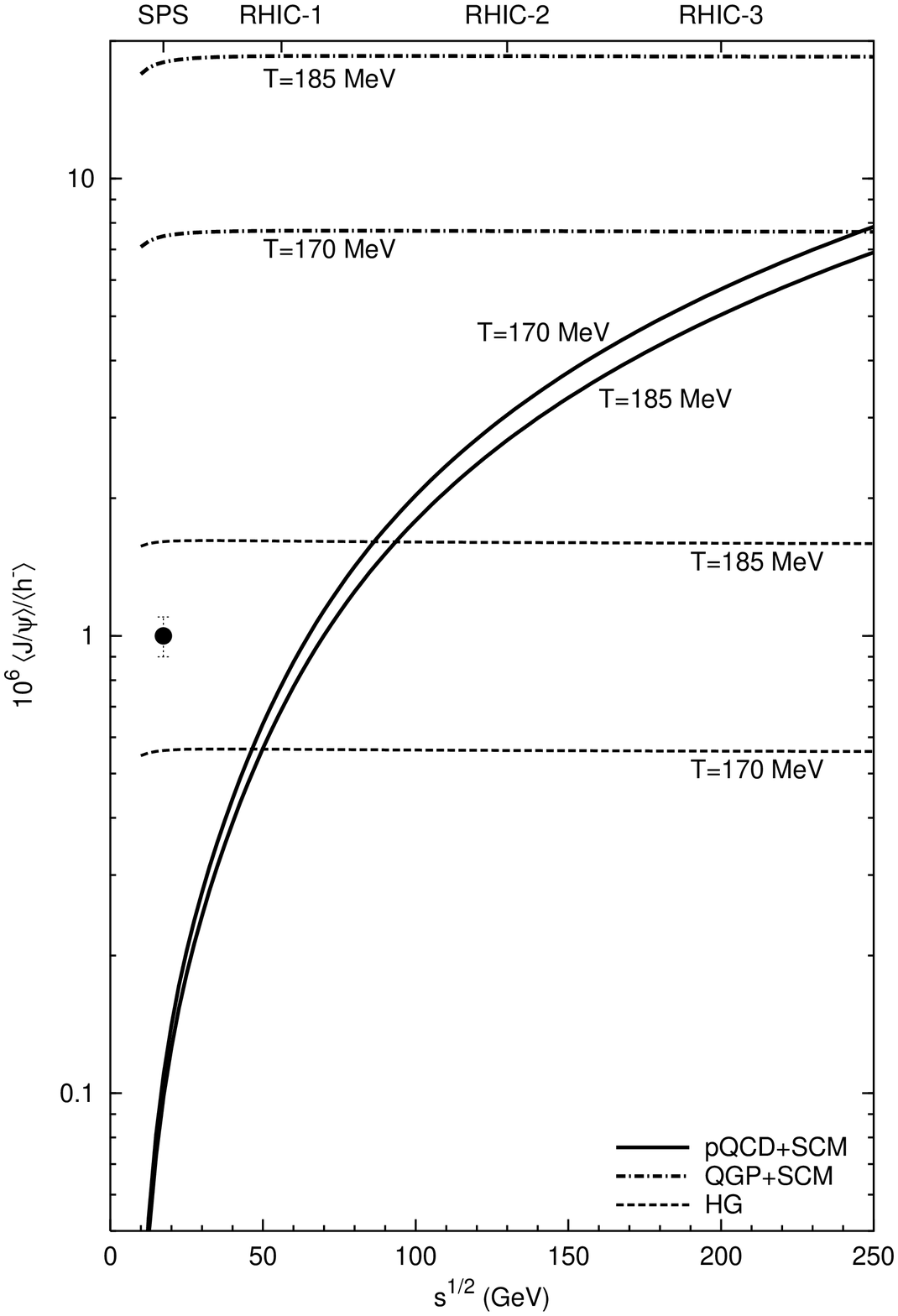}
\vfill
\mbox{}\\
\caption{\label{Jpsi_sl_hmi}
The dependence of the ratio of the $J/\psi$-multiplicity to
the negative hadron multiplicity
in Au+Au central collisions
on the center-of-mass energy of the nucleon pair.
The calculations are done in
the framework of three
different models:
statistical coalescence model with the number of
$c\overline{c}$ pairs given by perturbative QCD (pQCD+SCM), statistical
coalescence model with the number of $c\overline{c}$ pairs given by
equilibrium quark-gluon plasma (QGP+SCM) and equilibrium hadron gas model
(HG). All the dependences are shown
for two different temperatures: $T=170$ MeV and $T=185$ MeV.
The filled circle shows the value estimated from the NA50 experimental
data for Pb+Pb collisions at the SPS energy [2,14].
}
\end{center}
\end{figure}

The $J/\psi$ multiplicity is shown in Fig.~\ref{Jpsi}
and the ratio of $J/\psi$ to negative hadrons in Fig.~\ref{Jpsi_sl_hmi}.
The experimental estimate of the $J/\psi$ to $h^-$ ratio
extracted from the NA50 data in Pb+Pb collisions at 158~A~GeV
is also presented in Fig.~\ref{Jpsi_sl_hmi}.
It seems to be in agreement with
simple HG results \cite{Ga1} (i.e. $\gamma_c \approx 1$).
In the pQCD+SCM model the enhancement
of the total charm $N_{c\overline{c}}$ by a factor of about 3
is needed to explain the observed $J/\psi$
multiplicity \cite{Go:00}.
This prediction of the statistical coalescence model can be tested
in the near future (measurements of the open charm are planned
at CERN). Such a comparison will require to specify also
more  accurately the  $\langle J/\psi \rangle$ data.
A possible source of the open charm enhancement in A+A
collisions at the SPS with respect to the direct extrapolation
(\ref{pert}) from p+p to A+A may be the broadening of the phase
space available for the open charm production due to the presence
of the quark-gluon medium \cite{HFE}. This effect, however, is expected
to be small at RHIC energies.

\subsection{$J/\psi$ Multiplicity Ratio to Open Charm}

\begin{figure}[t]
\begin{center}
\vfill
\leavevmode
\epsfysize=17.5cm \epsfbox{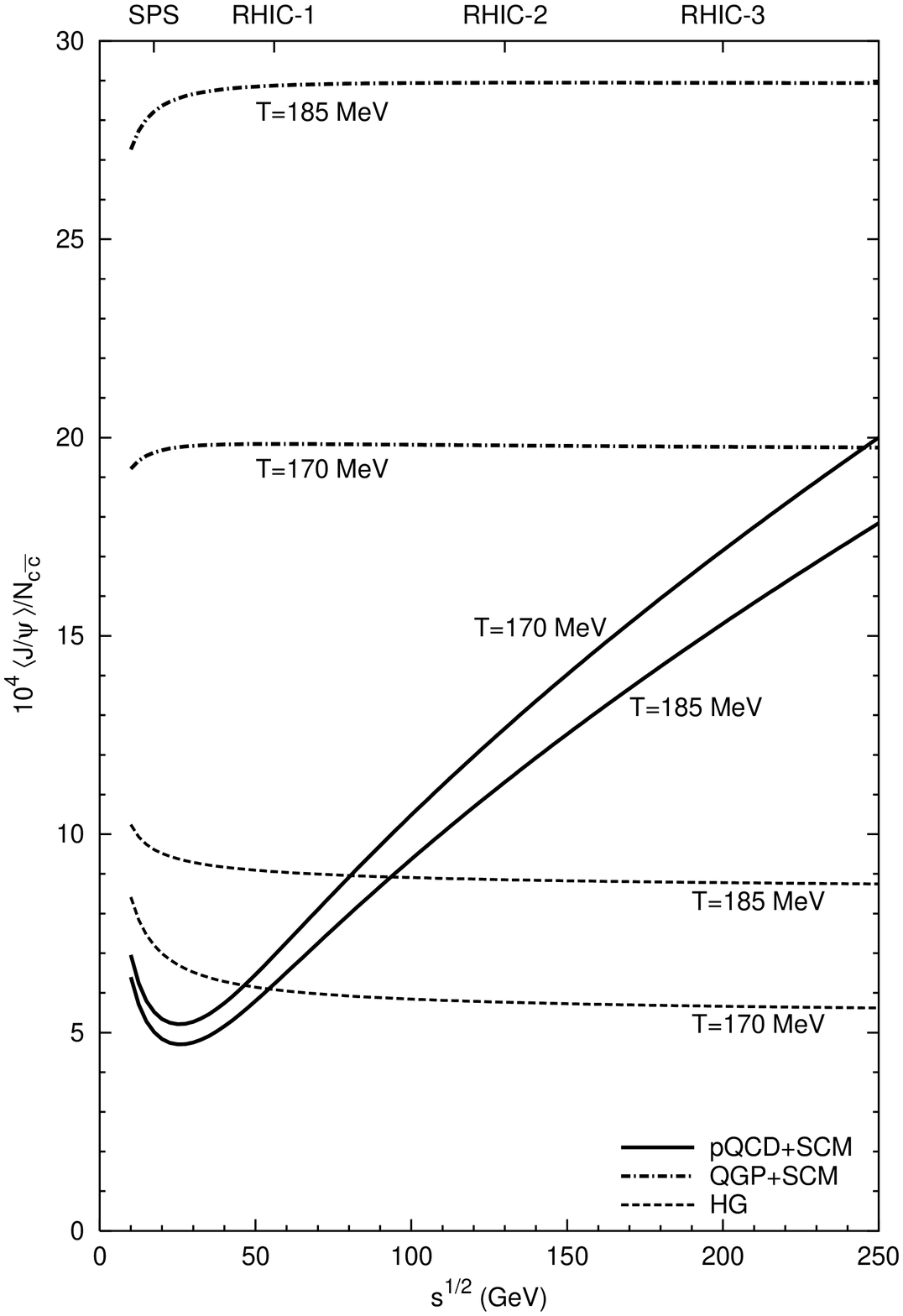}
\vfill
\mbox{}\\
\caption{The dependence of the ratio of the $J/\psi$-multiplicity to
the multiplicity of $c\overline{c}$ pairs
in Au+Au central collisions
on the center-of-mass energy of the nucleon pair.
The calculations are done in
the framework of three
different models:
statistical coalescence model with the number of
$c\overline{c}$ pairs given by perturbative QCD (pQCD+SCM), statistical
coalescence model with the number of $c\overline{c}$ pairs given by
equilibrium quark-gluon plasma (QGP+SCM) and equilibrium hadron gas model
(HG). All the dependences are shown
for two different temperatures: $T=170$ MeV and $T=185$ MeV.
\label{Jpsi_sl_ncc}
}
\end{center}
\end{figure}

The dependence of the $J/\psi$ to $N_{c\overline{c}}$ ratio
on the collision energy is shown in Fig.~\ref{Jpsi_sl_ncc}.
In the HG model and in QGP+SCM, the ratio of the $J/\psi$
to the number of $c\overline{c}$ pairs is constant at high
collision energies. This is because both the $J/\psi$ multiplicity
and the number of $c\overline{c}$ pairs are proportional
to the system volume $V$. In contrast, the pQCD+SCM model predicts
an essential increase of the ratio by a factor of about $3$
from the lowest to the highest RHIC energy.

\begin{figure}[t]
\begin{center}
\vfill
\leavevmode
\epsfysize=17.5cm \epsfbox{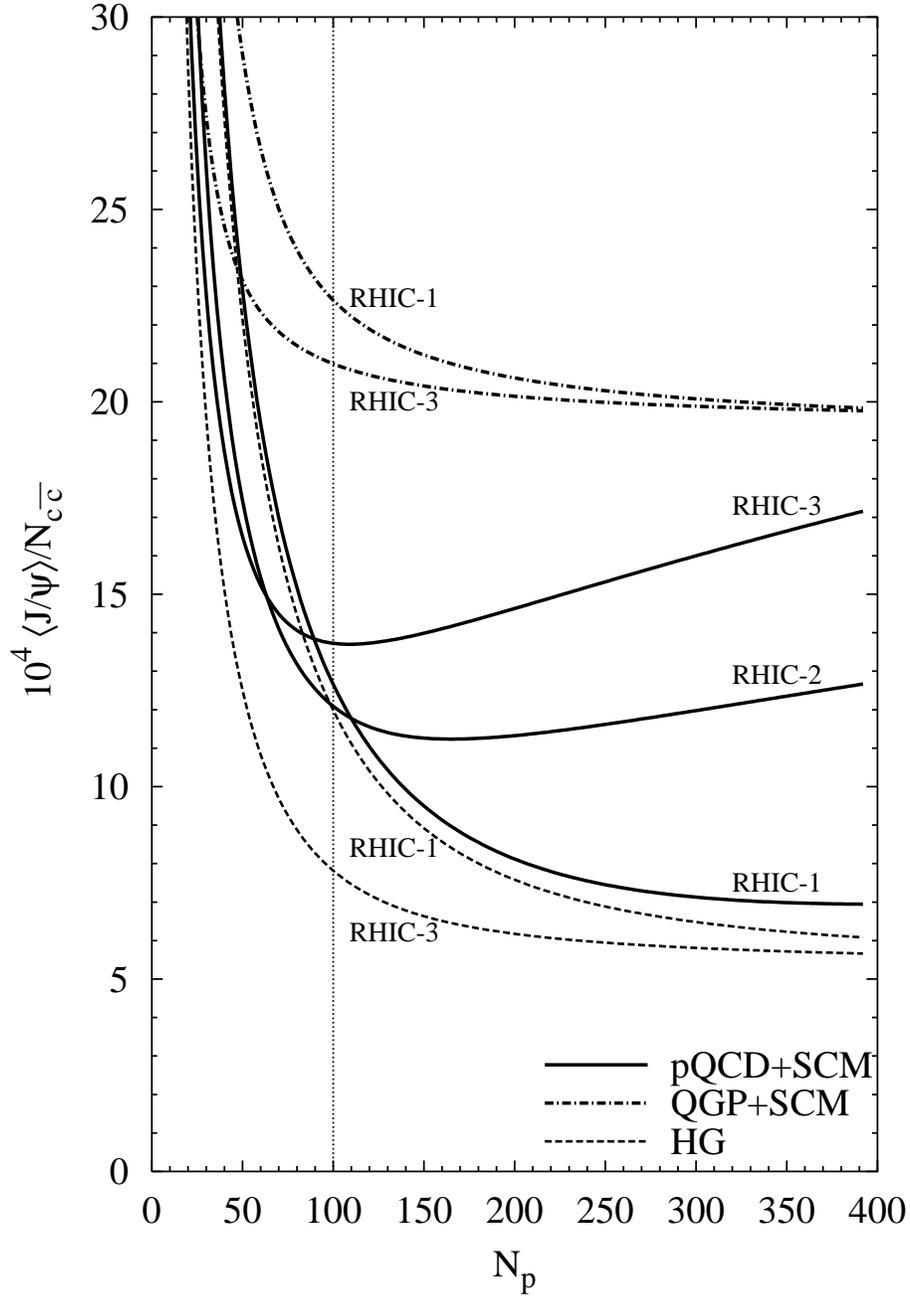}
\vfill
\mbox{}\\
\caption{The dependence of the ratio of the $J/\psi$-multiplicity to
the multiplicity of $c\overline{c}$ pairs on the number of nucleon
participants $N_p$ for Au+Au collisions.
The calculations are done in
the framework of three
different models:
statistical coalescence model with the number of
$c\overline{c}$ pairs given by perturbative QCD (pQCD+SCM), statistical
coalescence model with the number of $c\overline{c}$ pairs given by
equilibrium quark-gluon plasma (QGP+SCM) and equilibrium hadron gas model
(HG). In pQCD+SCM case, the dependences are shown for three
collision energies per the nucleon pair:
RHIC-1 ($\sqrt{s}=56$~GeV), RHIC-2 ($\sqrt{s}=130$~GeV)
and RHIC-3  ($\sqrt{s}=200$~GeV). For QGP+SCM and HG, the results at
RHIC-1 and RHIC-3 are presented. The freeze-out temperature was fixed
at $T=170$ MeV. The models are assumed to be valid for $N_p \ge 100$,
i.e. to the right from the dotted vertical line.
\label{Np_dep}}
\end{center}
\end{figure}

Results presented in Figs.~\ref{Ncc}--\ref{Jpsi_sl_ncc}
correspond to the most
central Au+Au collisions ($N_p\approx 2A$).
The dependence of the ratio $R$ on the number of
participating nucleons $N_p$ is shown in Fig.~\ref{Np_dep}.
(Note that statistical models are expected to be valid
for $N_p > 100$.)
In the pQCD+SCM dependencies,
one observes both the {\it $J/\psi$ suppression} and
the {\it $J/\psi$ enhancement} effects: the ratio (\ref{ratio})
decreases with $N_p$ when the number of created $c\overline{c}$ pairs
is small, $N^{pQCD}_{c\overline{c}} \alt 1$, but it begins
to increase with $N_p$ when
 $N^{pQCD}_{c\overline{c}}$ is essentially larger than unity.
In the HG model and in QGP+SCM, the suppression is also observed at small
$N_p$ corresponding to $N^{pQCD}_{c\overline{c}} \alt 1$. In contrast to
pQCD+SCM, the models with statistical charm production do not demonstrate
$J/\psi$ enhancement: at large $N_p$, which correspond to
$N^{pQCD}_{c\overline{c}} >> 1$, the ratio $R$ tends to a constant value.

\subsection{Centrality dependence of $\gamma_c$}

\begin{figure}[t]
\begin{center}
\vfill
\leavevmode
\epsfysize=17.0cm \epsfbox{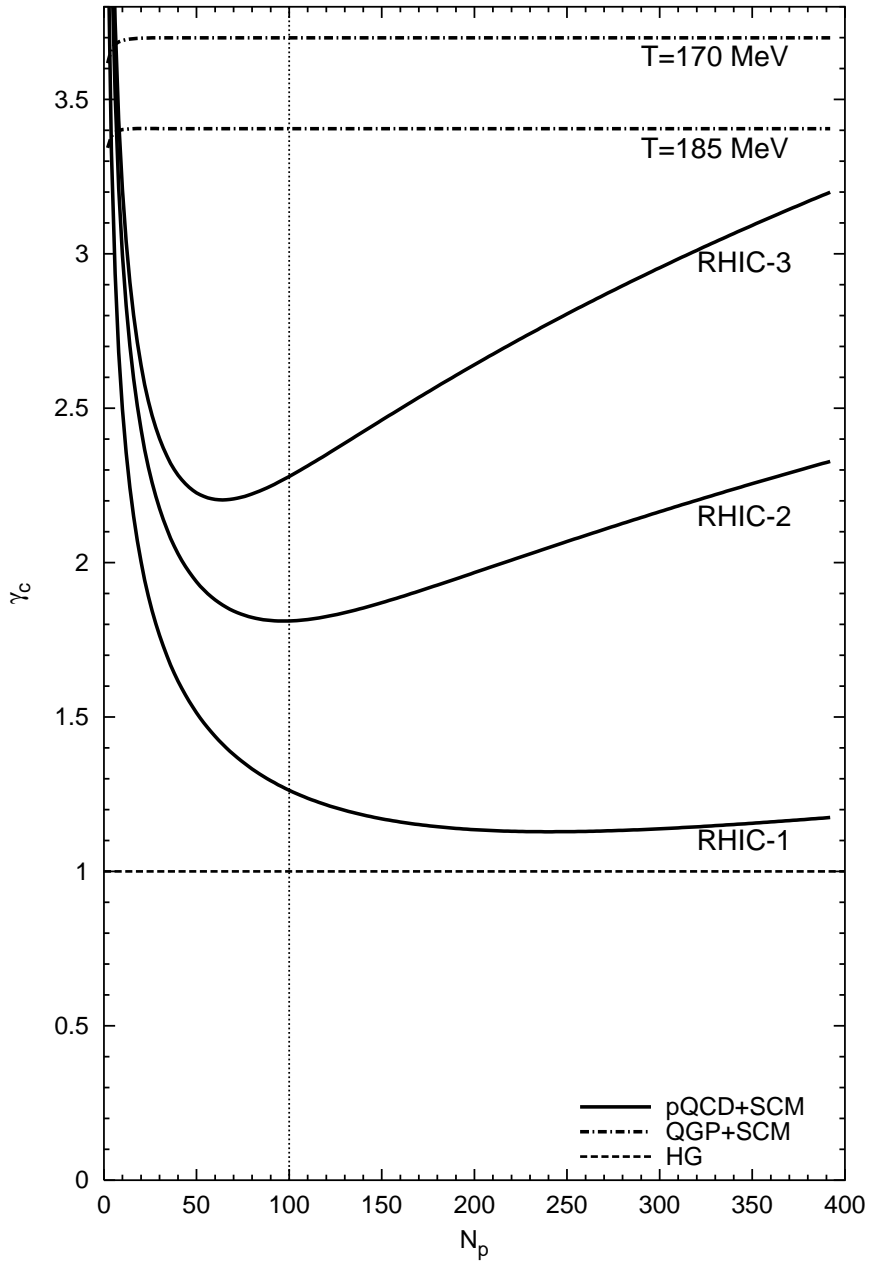}
\vfill
\mbox{}\\
\caption{The dependence of the charm enhancement factor $\gamma_c$ on 
the number of nucleon participants $N_p$ for Au+Au collisions.
The results are presented for three
different models:
statistical coalescence model with the number of
$c\overline{c}$ pairs given by perturbative QCD (pQCD+SCM), statistical
coalescence model with the number of $c\overline{c}$ pairs given by
equilibrium quark-gluon plasma (QGP+SCM) and equilibrium hadron gas model
(HG). In pQCD+SCM case, the dependences are shown for three
collision energies per the nucleon pair:
RHIC-1 ($\sqrt{s}=56$~GeV), RHIC-2 ($\sqrt{s}=130$~GeV)
and RHIC-3  ($\sqrt{s}=200$~GeV). 
The freeze-out temperature was fixed
at $T=170$ MeV. For QGP+SCM, the value of $\gamma_c$  is shown at two values of 
freeze-out temperature $T=170$ MeV and $T=185$ MeV (the dependence on the 
collision energy is negligeble). 
For the equilibrium hadron gas model
(HG) $\gamma_c \equiv 1$ by the definition.
The models are assumed to be valid for $N_p \ge 100$,
i.e. to the right from the dotted vertical line.
\label{gammac_Np_dep}}
\end{center}
\end{figure}

The dependence of charm enhancement factor $\gamma_c$ on the number of
participating nucleons $N_p$ is shown in Fig.~\ref{gammac_Np_dep}.
In the case of pQCD+SCM model, the behavior is similar to that of ratio $R$: 
$\gamma_c$ decreases with $N_p$ provided that the number of
$c\overline{c}$ pairs is small $N^{pQCD}_{c\overline{c}} \alt 1$ and 
increases at large $N^{pQCD}_{c\overline{c}}$. In contrast, the QGP+SCM
model leads to a nearly constant value of $\gamma_c$, which can be easily explaned 
compareing Eqs.(\ref{Ncc1}) and  (\ref{NQce}) and neglecting the small term
$\gamma_c^2 N_H$. From the same reason the dependence of $\gamma_c$ on $\sqrt{s}$
is also negligeble. (Let us remind that in the case of HG model, $\gamma_c \equiv 1$
by the definition).

\subsection{Discussion}

The results presented in Figs.~\ref{Jpsi_sl_ncc}--\ref{gammac_Np_dep}
dependences can be easily studied analytically in the limiting cases. For
$N_{c\overline{c}} >> 1$ one finds from Eq.(\ref{Ncc1}): 
\begin{equation}\label{ratio1}
R=
\frac{\langle J/\psi \rangle} {N_{c\overline{c}}}~ \sim~ \gamma_c ~\approx~
2N_{c\overline{c}}/N_O ~\sim~
\frac{N_{c\overline{c}}}{V} ~\sim~
\frac{N_{c\overline{c}}}{\langle \pi
\rangle}. 
\end{equation}
where $\langle \pi \rangle$ is the total pion multiplicity.
Therefore, in the limiting case $N_{c\overline{c}} >> 1$,
the ratio of the $J/\psi$ to the open charm (or a probability
to form the charmonium states) is proportional to the
factor $\gamma_c$, which in its turn is proportional 
to the density
of $c\overline{c}$ pairs.
In the case of pQCD+SCM, this $c\overline{c}$ density
increases both with collision energy $\sqrt{s}$ and with the number of
nucleon participants $N_p$:\\
1) the increase (\ref{pert}) of
$N^{pQCD}_{c\overline{c}}$ with $\sqrt{s}$ is much stronger
than the increase of the system volume $V$ or, equivalently,
the increase of the total pion multiplicity (\ref{pionexp});\\
2) the increase of $N^{pQCD}_{c\overline{c}}\sim N_p^{4/3}$
is stronger than $V\sim \langle \pi \rangle$ which is approximately
proportional to $N_p$.

In the case of thermal charm production (QGP+SCM and HG), the number
of $c\overline{c}$ pairs is proportional to the system volume.
Therefore the ratio R becomes constant when $N_{c\overline{c}} >> 1$.

The behavior of the $\langle J/\psi \rangle /
 N_{c\overline{c}} $ ratio is rather different
when $N_{c\overline{c}} << 1$ and the canonical suppression
factor in Eq.(\ref{Ncc1}) plays a crucial role. One finds
$\gamma_c^2 \approx 4 N_{c\overline{c}}/N_O^2$ and then
$\gamma_c \sim N_P^{-1/3}$ for pQCD+SCM.
The value of $R$, however, does not depend on $\gamma_c$:
\begin{equation}\label{ratio2}
R = \frac{\langle J/\psi \rangle} { N_{c\overline{c}}}~ \sim~
\frac{1}{V} ~\sim~
\frac{1}{\langle \pi \rangle} ~.
\end{equation}
The ratio (\ref{ratio})
decreases with both the collision energy
(approximately like $(\sqrt{s})^{1/2}$)
and the number of nucleon participants
(approximately like $N_p^{-1}$) and does not depend on the charm
production mechanism. This behavior is similar to the
standard picture of the $J/\psi$ suppression.

The {\it $J/\psi$ suppression} in pQCD+SCM happens at the SPS energy.
This energy is still "low" as $N^{pQCD}_{c\overline{c}} <1$
even in the most central Pb+Pb collisions.
The {\it $J/\psi$ suppression} (i.e. the
decrease of the $J/\psi/N^{pQCD}_{c\overline{c}}$  ratio)
with $N_p$ and/or $\sqrt{s}$
continues up to the RHIC energies  for the
peripheral collisions
with small number of nucleon participants.
This changes into the  {\it $J/\psi$ enhancement}
when both $N_p$ and  $\sqrt{s}$ are "large".
Therefore, if the $c\overline{c}$ production is described
by the pQCD, the statistical coalescence model
predicts the decrease of the ratio $R$ (\ref{ratio})
with $\sqrt{s}$ and/or $N_p$ (the $J/\psi$ {\it suppression})
when $N^{pQCD}_{c\overline{c}}  << 1$.
However, this ratio increases
with $\sqrt{s}$ and/or $N_p$ (the $J/\psi$ {\it enhancement})
when $N^{pQCD}_{c\overline{c}} >> 1$.
These limiting behaviors are smoothly connected
in the intermediate region of $N^{pQCD}_{c\overline{c}}\simeq 1$.

Other models for $J/\psi$ production via coalescence of
$c$ and $\overline{c}$ at the hadronization stage \cite{Ra:00,Le:00} give
similar {\it qualitative} result: the $J/\psi$ {\it enhancement} for
central collisions at high energies. {\it Quantitative} predictions,
however, are rather different.
In Ref.\cite{Ra:00}
$\langle J/\psi \rangle = (2.0\div 3.0)~10^{-1}$
per central Au+Au event
at $\sqrt{s}=200$~GeV, whereas our result in pQCD+SCM is
$\langle J/\psi \rangle \simeq 1.0~10^{-2}$
(note that $N_{c\overline{c}}$ estimates are very close:
approximately $10 \div 11$ $c\overline{c}$ pairs per event).
The predictions of the microscopical coalescence  model
of Ref.\cite{Le:00}
can be compared with our results at $\sqrt{s}= 130$~GeV.
They are rather close for  $J/\psi$ to negative hadron ratio:
$\langle J/\psi \rangle / \langle h^- \rangle =1.6 \ 10^{-6}$
\cite{Le:00} and
$\langle J/\psi \rangle / \langle h^- \rangle \simeq 2~10^{-6}$
in the pQCD+SCM. However, the total charm production is rather different:
$N_{c\overline{c}} \approx 12$ in
Ref.\cite{Le:00} and $N_{c\overline{c}} \approx 5$ in our
pQCD estimation.

\section{Conclusion}

The statistical production of $J/\psi$ mesons
in Au+Au collisions at RHIC energies has been
studied.
The obtained results are qualitatively different from
the standard picture of the $J/\psi$ suppression in $A+A$ collisions
\cite{Satz1,Satzr}, if the average number of produced charmed
quark-antiquark pairs per event is large
$N_{c\overline{c}} >> 1$. The standard
picture leads to monotonic
decrease of the ratio $R$ (the ratio of the $J/\psi$ multiplicity to
the total number of charm quark-antiquark pairs) with both the collision
energy and the number of participating nucleons. We predict
another behavior. The reason for this difference is the completely
distinct physical pictures for charmonia formation. The
standard picture supposes the charmonia states to be created
{\it exclusively} in primary nucleon-nucleon collisions
at the early stage of the reaction so that all subsequent interactions
can only destroy them.
We study another possibility: creation of charmed and charmonium
particles at the
hadronization stage.
In our opinion, the most probable scenario of the charmonium
production in A+A collisions at the RHIC energies is the following
(we call it pQCD+SCM): $c\overline{c}$ pairs are produced
exclusively in hard
parton collisions at the early stage of A+A reaction (pQCD), and
these charmed quarks and antiquarks are redistributed in the
open and hidden charm hadrons according to the statistical
mechanics prescriptions at the hadronization stage (SCM).
This model predicts a strong increase of the ratio $R$ with the collision
energy or the number of nucleon participants
($J/\psi$ {\it enhancement}), provided that the total
number of produced $c\overline{c}$ pairs per event is
essentially larger than $1$. This condition is fulfilled at the
highest RHIC energy $\sqrt{s}=200$~GeV already for $N_p=100$
and also at lower RHIC energies
for the most central Au+Au collisions ($N_p\simeq 2A$).

To get a feeling of possible effects of the charm creation
(annihilation) by secondary parton and/or hadron rescatterings we
have considered
two extreme models of charm chemical equilibration: in the quark gluon
plasma (QGP+SCM) and in the hadron gas (HG).
Both models predict almost constant
ratio $R$ at $N_{c\overline{c}} >> 1$. There is however
a quantitative difference: QGP+SCM predicts much larger value of $R$
and much larger value of the total charm than pQCD+SCM and
HG models. The predictions of the HG model for the open charm and
charmonium production at the upper
RHIC energies are below the results of the pQCD+SCM and QGP+SCM.

Our predictions for the behavior of $R$ at $N_{c\overline{c}} >> 1$
($J/\psi$ enhancement) in the pQCD+SCM are in
a qualitative agreement with
other coalescence models
\cite{Ra:00,Le:00}, but quantitatively the results are rather
different.

We conclude that $Au+Au$ collision experiments at RHIC offer a
possibility to disentangle various models of the open charm and
charmonia production. Even those models that give very similar
results at SPS energies become quantitatively and even qualitatively
different at RHIC. Measurements of the $J/\psi$ productions combined with
the open charm measurements are needed to give a definite answer
about the dominant mechanisms of hidden and open charm production
in ultrarelativistic collisions of heavy nuclei.

\acknowledgments
The authors
are thankful to F. Becattini, P. Braun-Munzinger, K.A.~Bugaev,
M. Ga\'zdzicki,
L. Gerland,
I.N.~Mishustin, G.C.~Nayak,  K.~Redlich and J.~Stachel for comments and
discussions.
We acknowledge the financial support of GSI and DAAD,
 Germany.
The research described in this publication was made possible in part by
Award \# UP1-2119 of the U.S. Civilian Research and Development
Foundation for the Independent States of the Former Soviet Union
(CRDF). This manuscript has been authorized under Contract
No. DE-AC02-98H10886 with
the U. S. Department of Energy.

\begin{table}
\vspace{0.5cm}
\caption{\label{table1} }
\vspace{0.5cm}
\begin{tabular}{|c||c|c|c|}
\hline
\rule[-4mm]{0mm}{10mm}
&\hspace{1.2cm} {\bf pQCD + SCM}  \hspace{1.2cm}
&\hspace{1.2cm} {\bf QGP + SCM} \hspace{1.2cm}
&\hspace{1.4cm} {\bf HG}         \hspace{1.4cm}   \\
\hline \hline
\rule[-4mm]{0mm}{10mm}
$N_{c\overline{c}}$  & $N^{pQCD}_{c\overline{c}}$~,~
Eq.~(\ref{pert})&
$N^{QGP}_{c\overline{c}}$~,~Eq.~(\ref{pert}) &
$N^{HG}_{c\overline{c}}$~,~Eq.~(\ref{NccHG}) \\
\hline
\rule[-4mm]{0mm}{10mm}
$\gamma_c$ &
Eq.~(\ref{Ncc1})~,~$N_{c\overline{c}}=N^{pQCD}_{c\overline{c}}$ &
Eq.(\ref{Ncc1})~,~$N_{c\overline{c}}=N^{QGP}_{c\overline{c}}$ &
$\gamma_c\equiv 1$\\
\hline
\rule[-4mm]{0mm}{10mm}
$\langle J/\psi \rangle$ & $\gamma_c^2~N_{J/\psi}^{tot}$~, Eq.~(\ref{Npsi}) &
$\gamma_c^2~N_{J/\psi}^{tot}$~, Eq.~(\ref{Npsi}) &
$N_{J/\psi}^{tot}$ , Eq. (\ref{psitot})\\
\hline
\end{tabular}
\end{table}

\end{document}